\documentclass[11pt]{article}
\usepackage{graphicx}
\usepackage{amssymb,amsthm,amsmath,float}
\usepackage{epstopdf}
\newcommand{\Th}[1]{Th.~\ref{#1}}
\newcommand{\Lem}[1]{Lem.~\ref{#1}}

\newcommand{\Sec}[1]{\S \ref{#1}}
\newcommand{\Fig}[1]{Fig.~\ref{#1}}
\newcommand{\Tbl}[1]{Table~\ref{#1}}

\newcommand{\InsertFig}[4]
{\begin{figure}[ht]
       \centerline{
         \includegraphics[width=#4]{./#1}
       }
       \caption{{\footnotesize  #2}
       \label{#3}}
\end{figure}}

\newcommand{\R}{{\mathbb{ R}}}

\newcommand{\N}{{\mathbb{ N}}}

\newcommand{\cD}{{\cal D}}

\newcommand{\tr}{\mathop{\rm tr}}
\newcommand{\ad}{\mathop{\rm ad}\nolimits}
\newcommand{\Ad}{\mathop{\rm Ad}\nolimits}
\newcommand{\diag}{\mathop{\rm diag}}
\newcommand{\rng}{\mathop{\rm rng}}
\newcommand{\coker}{\mathop{\rm coker}}

\newcommand{\Pnd}{{P^n_d}}
\newcommand{\Fnd}{{F^n_d}}
\newcommand{\Vnd}{{V^n_d}}
\newcommand{\Und}{{U^n_d}}

\newtheorem{thm}{Theorem}

\newtheorem{lem}[thm]{Lemma}

\theoremstyle{definition}
\newtheorem*{example}{Example}
\theoremstyle{remark}
\newtheorem{rem}{Remark}

\newcommand{\Lvf}{\mathcal{L}}
\newcommand{\Lmap}{\mathrm{L}}
\newcommand{\Id}{\mathbb{I}}
\renewcommand{\sl}{\mathfrak{sl}}
\renewcommand{\sp}{\mathfrak{sp}}
\newcommand{\so}{\mathfrak{so}}

\title{Nilpotent normal form for divergence-free vector fields and 
volume-preserving maps}
\author{  
        H.~R.~Dullin\thanks{H.R.Dullin@lboro.ac.uk;
        on leave from Loughborough University, UK;
        HRD was supported in part by a Leverhulme Research Fellowship,
        thanks to MSRI Berkeley and UC Boulder for their hospitality.} 
       ~and J.~D.~Meiss\thanks
      {
        James.Meiss@colorado.edu, 
        JDM was supported in part by NSF grant DMS-0202032 and
        the Mathematical Sciences Research Institute.
      }\\
     Department of Applied Mathematics\\
     University of Colorado \\
     Boulder, CO 80309-0526 \\
}
\date{\today}
\begin{document}
\maketitle

\begin{abstract}
\vspace*{1ex}
\noindent
We study the normal forms for incompressible flows and maps in the neighborhood of an equilibrium or fixed point with a triple eigenvalue. We prove that when a divergence free vector field in $\R^3$ 
has nilpotent linearization with maximal Jordan block then, to arbitrary degree, 
coordinates can be chosen so that the nonlinear terms occur as a single function of two variables in the 
third component. The analogue for volume-preserving diffeomorphisms gives 
an optimal normal form in which the truncation of the normal form at any degree
gives an exactly volume-preserving map whose inverse is also polynomial inverse with the same degree.
\end{abstract}

\section{Introduction}\label{sec:Introduction}
A system of ODEs $\dot{\xi} = v(\xi,t)$ gives rise to a volume-preserving flow when the vector field $v$ has zero divergence,
\begin{equation}\label{eq:divergence}
	\nabla \cdot v = 0 \;.L
\end{equation}
Similarly, a diffeomorphism $f: \R^n \to \R^n$ is volume preserving when its Jacobian has unit determinant, $\det{Df(\xi)} = 1$. Volume-preserving dynamics arises, for example, in the flow of a Lagrangian  element in an incompressible fluid \cite{Holmes84, Dombre86, Feingold88, Cartwright96, Shinbrot01, Speetjens04, Mullowney05, Mullowney06}, or the tracing of lines in a magnetic field, \cite{Thyagaraja85, Lau92, Greene93}. Thus such flows are of interest in the study of the motion of passive tracers in time-dependent, incompressible fluids and for the dynamics of charged particles in strong magnetic fields. Volume-preserving maps are a natural generalization of area-preserving maps to higher dimensions.  They also arise as the normal form for certain homoclinic bifurcations for three-dimensional systems \cite{Gonchenko05a, Gonchenko06}, and as integrators for incompressible flows \cite{Qin93, Quispel95, Kang95, Suris96} and thus have intrinsic mathematical interest \cite{Cheng90b, Baier90, Xia92, Romkedar93, Lomeli98a,Gomez02,Lomeli00b, Lomeli03}.

An equilibrium of a volume-preserving flow can undergo a number of codimension-one bifurcations. For example when one of the eigenvalues vanishes and the remaining eigenvalues do not lie on the imaginary axis, then the resulting bifurcation is generically of standard saddle-node type. Similarly, when a single pair of eigenvalues lie on the imaginary axis, the bifurcation is of Hopf type. Since the eigenvalues of the equilibrium satisfy $\sum_{i=1}^n \lambda_i = 0$, there are two more exotic cases that are codimension-one only for low dimensions. For three dimensions, the codimension-one configuration $\{ 0, i\omega , -i \omega\}$ can give rise, through a Hopf-saddle node bifurcation to the creation of a periodic orbit surrounded by a family of invariant tori \cite{Broer81}. The final codimension-one case occurs in four dimensions when the eigenvalues are $\{ i\omega_1, -i\omega_2, i\omega_2, -i\omega_2\}$; this bifurcation is analogous to the Hamiltonian-Hopf bifurcation \cite{Broer81}.

The linearization of a volume-preserving map at a fixed-point $x^* = f(x^*)$ gives a matrix $M = Df(x^*)$ with determinant one, so that the multipliers satisfy $\prod_{i=1}^d \lambda_i = 1$. For the three-dimensional case, the characteristic polynomial has the form
\begin{equation}\label{eq:characteristic}
    p(\lambda) = \det(\lambda I - M) = \lambda^3 -\tau \lambda^2 +\sigma \lambda -1 \;.
\end{equation}
where $\tau$ denotes the trace, $\tau = \tr{M}$, and $\sigma$ the second-trace, $\sigma = \frac12 \left( (\tr{M})^2 - \tr{M^2}\right)$. Consequently, $(\tau,\sigma)$ parametrize the space of volume-preserving matrices that are linearly conjugate to $M$, at least when the eigenvalues are distinct. The eigenvalue configurations in this space, see \Fig{fig:stability} \cite{Lomeli98a, Lomeli00b}, show that there are two codimension-two points. The first, at $(\tau,\sigma) = (-1,-1)$ where the eigenvalues are $(1,-1,-1)$ corresponds to simultaneous period-doubling and saddle-node bifurcations, and the second, at $(\tau,\sigma) = (3,3)$, to the triple eigenvalue $(1,1,1)$. These are connected by a line segment $\{\tau = \sigma : -1 < \tau < 3\}$ on which two eigenvalues are on the unit circle, and the third is equal to one, corresponding to simultaneous saddle-node and Neimark-Sacker bifurcations. 
\InsertFig{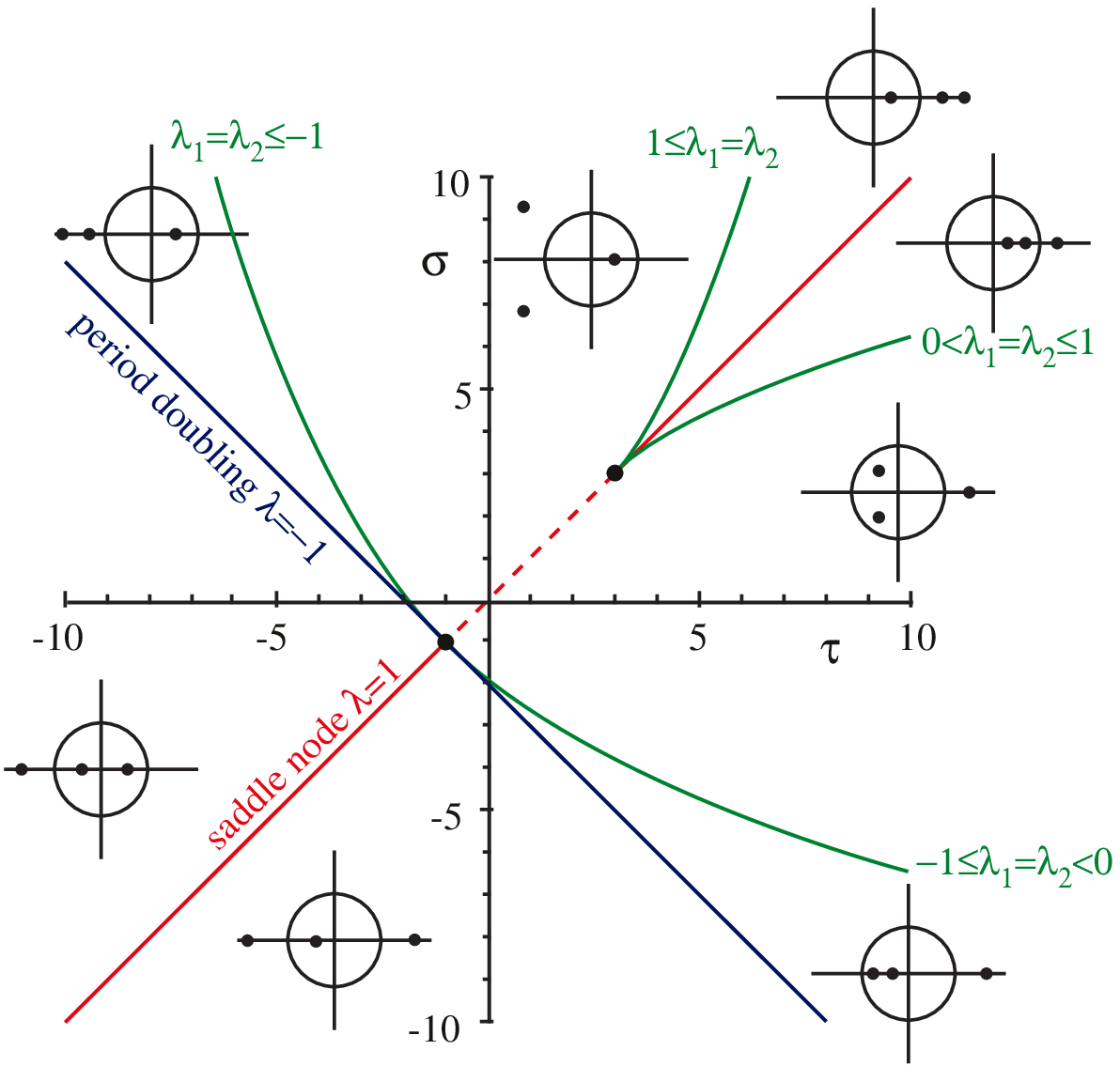}{Classification of the eigenvalues for a three-dimensional, 
volume-preserving map as a function of the trace $\tau$ and second trace $\sigma$.}{fig:stability}{3in}

In this paper we study the normal forms for the codimension-two cases of a three-dimensional incompressble flow with eigenvalues $\{0,0,0\}$ and of a volume-preserving map with multipliers $\{1,1,1\}$. We first recall some of the standard results on normal forms.

Near an equilibrium, a set of ODEs with a smooth vector field takes the form
\begin{equation}\label{eq:original}
   \dot{\xi} = v(\xi) = J \xi + b(\xi) \;,
\end{equation}
for  $\xi \in \R^n$, with a vector field $v$ whose linear part is $J \xi$ and nonlinear part is $b(\xi) = O(2)$. A local normal form is a conjugate system
\begin{equation}\label{eq:transformed}
   \dot{\eta} = w(\eta) = J \eta + c(\eta) \;,
\end{equation}
that is ``simpler" in some sense. For example, it is usually desirable to eliminate as many of the nonlinear terms in $b$ as possible since then the vector field $w$ will have fewer parameters. A first step in this process is to choose coordinates in which $J$ itself is simple, and it is typical to begin by normalizing the linear part, that is to choose $J$ to be in Jordan normal form.

Since the linear parts of  \eqref{eq:original} and  \eqref{eq:transformed} are the same, the transformation can be assumed---to lowest order---to be the identity,
\begin{equation}\label{eq:conjugacy}
	\eta = \psi(\xi) = \xi + h(\xi) \;.
\end{equation}
Though the ``simplest" form of a flow may only be topologically conjugate to the original system (as in the Hartman-Grobman theorem when $J$ is hyperbolic), it is not possible to explicitly construct $w$ unless we assume that both the vector fields and the transformation are smooth. When $\psi$ is a diffeomorphism then the vector fields of  \eqref{eq:original} and  \eqref{eq:transformed} are related by
\begin{align*}
	D\psi(\xi)v(\xi) &= w(\psi(\xi)) \quad \Rightarrow \\
	b(\xi) + Dh(\xi)J\xi + Dh(\xi) b(\xi)) &=  Jh(\xi) + c(J\xi + h(\xi))
\end{align*}
Under the assumption that $v,w,h \in C^\infty$, the normal form can be computed by power series expansion. One way to accomplish this is to transform the terms of each degree one at a time. When this normalization has been carried out for all terms through degree $d-1$, i.e., $v = w + O(d)$, we let $h(\xi)= O(d)$ in \eqref{eq:conjugacy}; these degree $d$ terms must then solve the ``homological equation"
\begin{equation}\label{eq:homoEq}
	\Lvf_J(h)(\xi) =c(\xi) - b(\xi) + O(d+1)\,.
\end{equation}
Here $\Lvf_J$ is the homological operator defined by 
\begin{equation}\label{eq:homoVF}
  \Lvf_J = \ad_J  \equiv [  J\xi, \cdot ] \;,
\end{equation}
where  $[,]$ the Lie bracket of vector fields: 
\begin{equation}\label{eq:adDefine}
	\ad_J h(\xi) = (J\xi \cdot \nabla) h(\xi) - (h(\xi)\cdot \nabla) J\xi
			= Dh(\xi) J\xi - J h(\xi) \;.
\end{equation}
The problem of constructing $h$ at degree $d$ then reduces to linear algebra on the finite dimensional space, $\Fnd$, of $n$-dimensional vectors of polynomials in $n$-variables of homogeneous degree-$d$. In this case, a solution $h \in \Fnd$ to \eqref{eq:homoVF} exists providing $c-b \in \rng{\Lvf_J}\cap \Fnd$. If the operator $\Lvf_J$ were surjective on $\Fnd$, then we could set $c = 0$ to eliminate all nonlinear terms. However, the homological operator invariably has a nontrivial kernel, and $c$ must be chosen to eliminate any terms in $b$ that are not in $\rng{\Lvf_J}$.

Recall that any matrix $J = S+N$ has a unique splitting into commuting matrices such that $S$ is semisimple and $N$ is nilpotent. Moreover, $\ker J = \ker S \cap \ker N$. It can be similarly shown that $\Lvf_J = \Lvf_S + \Lvf_N$ is a semisimple-nilpotent splitting of the homological operator on $\Fnd$. The construction of a normal form requires the selection of a complement to $\rng{\Lvf_J}$; this can be chosen to be the intersection of complements to the ranges of the semisimple and nilpotent parts.

For the semisimple case, it is easy to find a basis for $\Fnd$ in which $\Lvf_S$ is also diagonal (the vector monomials, see \Sec{sec:gen}). In this case $\ker{\Lvf_S}$ is a complement to $\rng{\Lvf_S}$, so that $c$ can be chosen to be the terms in $b$ that correspond to eigenfunctions of the homological operator with zero eigenvalue. In this way the construction for the semisimple case reduces to characterizing $\ker{\Lvf_S}$.

The construction of a complement to the range of the nilpotent homological operator for a nilpotent matrix $N$ is not as easy. There are two commonly used methods: 
\begin{itemize}
	\item Given any scalar product $\langle, \rangle$ on the space of polynomial 
vector fields, then an orthogonal complement to the range is $\coker{\Lvf_N}=\ker{(\Lvf_N)^*}$, the kernel of the adjoint. A nice choice of scalar product, generalizing the Frobenius inner product of matrices, leads to $(\Lvf_N)^* = \Lvf_{N^*}$ \cite{Belitskii02,Elphick87}. This is called inner-product style in \cite{Murdock03}.
	\item From the representation theory of $\sl(2)$ and the Jacobson-Morozov 
embedding theorem, for any nilpotent $N$ there are matrices $M$ and $T$ 
that together give a representation of $\sl(2)$. It then follows that 
$\ker \Lvf_M$ is a complement to the range of $\Lvf_N$
\cite{CushmanSanders86, CushmanSanders88}. This is called $\sl(2)$ style in \cite{Murdock03}.
\end{itemize}

In general these approaches lead to different complements. Unfortunately, neither of these complements typically has the ``simplest" form. For example, one could declare the form with the minimal number of nonlinear terms at each degree to be simplest. Murdock as discussed a separate procedure that can be appended to either style to simplify the normal form in this way \cite{Murdock03}.

In this paper we study an incompressible vector field in three dimensions with a multiplicity-three eigenvalue $\lambda =0$ so that $J = 0 + N$. Generically $\lambda$ has geometric multiplicity one, so $N$ has a single Jordan block. We will show that the complement to $\rng{\Lvf}$ can be selected so that the normal form is particularly simple:

\begin{thm} \label{thm:VF}
Consider a smooth vector field  \eqref{eq:original} on $\R^3$, \
$\xi = (x,y,z)^T \in \R^3$, with vanishing divergence,  \eqref{eq:divergence} 
and linear part 
\begin{equation}\label{eq:3DN}
 Dv(0) = N = \begin{pmatrix} 0 & 1 & 0 \\ 0 & 0 & 1 \\ 0 & 0 & 0 \end{pmatrix} \,.
\end{equation}
The vector field $V$ can be transformed by a 
{\em volume-preserving},  near-identity transformation  \eqref{eq:conjugacy} into the form 
\begin{equation}\label{eq:flowNF}
	N\xi + \hat{e}_3 p(x, y), \quad \hat{e}_3 = (0,0,1)^T \,.
\end{equation}
up to terms of arbitrary degree. 
\end{thm}

The simple complement proposed here extends Murdock's results 
to the case of divergence-free vector fields. For the particular 
$N$ we are considering, the general result was already obtained in 
\cite{Elphick87}. It states that the nonlinear part of the transformed 
vector field has the form $\hat{e}_3 ( z \varphi_1 + y \varphi_2 + \varphi_3)$ 
where $\varphi_i = \varphi_i(x, y^2 - 2 x z)$ are arbitrary {\em polynomials}.
We will reconstruct this result in \Sec{sec:VectorField}, since we need
an intermediate form from this construction to obtain \Th{thm:VF}.
Indeed, as we explain in \Sec{sec:DivergenceFree}, it is not possible 
to simply impose zero divergence on this simplified form. In other words, 
the operations of imposing zero divergence and computing the simplified 
normal form do not commute.

A similar theorem holds for dimension two, i.e., the divergence-free case 
of the Takens-Bogdanov bifurcation. Generalization to four or more dimensions 
is not as straightforward, as we discuss in \Sec{sec:conclusion}.

Normal forms for maps can be found by analogous means. Let  $f : \R^n \to \R^n$
be a smooth diffeomorphism
\begin{equation}\label{eq:mapOriginal}
      f(\xi) = J(\xi + b(\xi))
\end{equation}
where $b = O(2)$ represents the nonlinear terms. A formal conjugacy 
\eqref{eq:conjugacy} $\psi \circ f = g \circ \psi$ to a normal form 
$g(\eta) = J(\eta+c(\eta))$ can be found if we can solve the homological equation
$ \Lmap_J h(\xi)   = c(\xi) - b(\xi)$ at each degree for $h$, where
\begin{equation}\label{eq:homoMap}
	\Lmap_J  \equiv \Ad_{J^{-1}}  - id
\end{equation} and $Ad_J h(\xi) \equiv J h( J^{-1}\xi)$. The terms 
$c$ in the normal form should again be selected to be in a 
complement to the range of the homological operator. Construction 
of this complement depends upon the properties of $J$.

Here we consider the case of volume-preserving maps that have a fixed point with multipliers $\lambda = 1$ of multiplicity three.  The mapping analogue of \Th{thm:VF} is

\begin{thm} \label{thm:MAP}
Consider a smooth, volume-preserving diffeomorphism  \eqref{eq:mapOriginal} on $ \R^3$ 
with the linear part
\begin{equation}\label{eq:MapNF}
 Df(0) = J = \begin{pmatrix} 1 & 1 & 0 \\ 0 & 1 & 1 \\ 0 & 0 & 1 \end{pmatrix}  = \Id + N \,.
\end{equation}
By a {\em volume-preserving}, near-identity transformation 
the map $f$ can be put into the form
\begin{equation}\label{eq:mapNF}
g(\xi) = J(\xi + \hat{e}_3 p(x, y)), \quad \hat{e}_3 = (0,0,1)^T.
\end{equation}
up to terms of arbitrary high degree. The truncation of the normal form at any 
degree is {\em exactly} volume preserving with a polynomial inverse of the same degree.
\end{thm}

The essential step for the transition from vector fields to maps is to use the 
exponential of the vector field and the exponential of $\ad$ instead of $\Ad$. 
However, since the simplified normal form style
is basis dependent, additional coordinate transformations are needed to 
transfer the result. We are motivated by the approach of \cite{BridgesCushman93}.

These results can be extended to the (formal) unfolding of these bifurcations,
as stated in 
\begin{thm} \label{thm:UF}
The unfoldings of the vector field of theorem~1 and the map of theorem~2 are obtained
by replacing the polynomial $p(x,y)$ in  \eqref{eq:flowNF} and  \eqref{eq:mapNF} by 
\[
\epsilon + \mu_1 x + \mu_2 y + p(x,y; \epsilon, \mu_1, \mu_2) \;, 
\]
where the lowest order terms of $p$ in $x$ and $y$ are quadratic.
\end{thm}

\section{Polynomial Vector Fields}\label{sec:gen}

Normal form theory uses formal power series of vector fields. In this section we introduce our notation for these vector fields and give some of the results about the polynomial subspaces and their bases that we will use.

For each point $\xi \in \R^n$ and each $m \in \N^n$, let $\xi^m = \xi_1^{m_1}\xi_2^{m_2} \dots \xi_n^{m_n}$ denote a scalar monomial with degree $|m| = \sum_{i=1}^n m_i$. Since the homological operators $\Lvf$  \eqref{eq:homoVF} and $\Lmap$  \eqref{eq:homoMap} preserve the degree, one can restrict to the finite-dimensional spaces of polynomial vector fields with fixed degree.
 
Let $P^n$ be the space of formal (without regard to convergence) polynomials in $n$-variables and  $\Pnd \subset P^n$ be those with homogeneous degree $d$, i.e., if $p 
\in \Pnd$, then $p(a\xi) = a^d p(\xi)$ for all $a\in \R$. One basis for $\Pnd$ is the set of all monomials $\{ \xi^m \,:\, |m| = d \}$. Consequently the dimension of $\Pnd$ is
\[
	\dim \Pnd = \binom{n+d-1}{d}  = \frac{(n+d-1)!}{(n-1)!d!} \;.
\]
The vector space of polynomial vector fields in $\R^n$ is denoted $F^n$ and the subspace with entries from $\Pnd$ in each component
is denoted
\begin{equation}\label{eq:FndDefine}
	 \Fnd =\{ h(\xi): h_i(\xi) \in \Pnd, i =1, \ldots, n\} \;.
\end{equation}
 A basis for this space is given by all ``vector monomials,''
\begin{equation}\label{eq:monomialBasis}
   p_{i,m} =  \xi^m \hat{e}_i = (0,\ldots, 0, \xi^m, 0,\ldots,0)^T \;, \; |m| = d, \; 1 \le i \le n \;,
\end{equation}
consequently,
\[
 \dim \Fnd = n \dim \Pnd \,.
\]
For example, $\dim P_d^2 = d+1$ and $\dim F_d^3 = 3 (d+2)(d+1)/2$.
It is usually obvious that the operators we consider that act on $F^n$ have 
$\Fnd$ as invariant subspaces, i.e., they preserve the degree. 

The space of divergence free vector fields of degree-$d$ is denoted by $\Vnd$:
\begin{equation}\label{eq:VndDefine}
    \Vnd = \{ v \in \Fnd \,:\, \nabla \cdot v  = 0 \} \,.
\end{equation}
Since $\nabla \cdot$ is a map from $\Fnd$ to $P_{d-1}^n$ the condition 
of vanishing divergence lowers the dimension by $\dim P_{d-1}^n$, and hence
\begin{equation}\label{eq:dimenVnd}
  \dim \Vnd = \dim \Fnd - \dim P_{d-1}^n \,.
\end{equation}
A basis for $ \Vnd$ can be constructed as follows:
\begin{lem}
	A basis of $ \Vnd$ consists of
	$n \dim P_d^{n-1}$ basis vectors $p_{i,\check{m}}$ where the $\check{}$ indicates
	that  $\check{m}_i = 0$, $|\check{m}| = d$, and 
	$(n-1)\dim P_{d-1}^{n}$ basis vectors of the form
	\begin{equation}\label{eq:vpbasis}
	    v_{i,m} =  
	      \xi^m \left[(1+m_{i+1})\xi_i \hat{e}_i -(1+m_i)\xi_{i+1} \hat{e}_{i+1}\right] \;,\; |m| = d-1 \;,
	\end{equation}
	with $1 \le i \le n-1$.
\end{lem}

\begin{proof}
First note that the dimensions of the two subsets add up:
\[
	\dim  \Vnd = \dim \Fnd - \dim P_{d-1}^n = n \dim P_d^{n-1} + (n-1) \dim P_{d-1}^{n} \,.
\]   
The first set of basis vectors has vanishing divergence since
\[
   \nabla \cdot p_{i, m} = \partial_i \xi^m = m_i \; \xi_1^{m_1} \dots \xi_i^{m_i - 1} \dots \xi_n^{m_n} \,,
\]
vanishes when $m_i = 0$. There are $n \dim P_d^{n-1}$ vector monomials of this type,
since in the $i$th component of the vector we can put a polynomial depending on all $n-1$ 
variables except $\xi_i$, and the dimension of this space is $\dim P_d^{n-1}$.
The remaining basis elements are vector fields of the form $A \xi$ multiplied by 
$\xi^m$ with $|m| = d-1$ and the diagonal matrix $A = \diag(a_1,\ldots,a_n)$. They are divergence free if
\[
    \nabla \cdot (\xi^m A\xi) = \sum_{i=1}^n \partial_i (\xi^m a_i \xi_i) = 
    \sum_{i=1}^n a_i(\xi_i \partial_i \xi^m +  \xi^m) =
    \xi^m \sum_{i=1}^n a_i  (m_i + 1) = 0 \;.
\]
For given $m$
this equation has $n-1$ independent solutions, those vectors that are orthogonal to the vector $( 1 + m_i)_{i=1,\dots, n}$. 
Hence there are $(n-1) \dim P_{d-1}^n$ such vector fields. 
The independent solutions can be chosen such that only two $a_i$ are nonzero, as in  \eqref{eq:vpbasis}. The $A \xi$ term gives monomials that depend on $\xi_i$ in component $i$, and are thus independent of the first group; they are also independent of each other because the monomials $\xi^m$ are distinct.
\end{proof}

There is a natural inner product on the space of polynomial 
vector fields that generalizes the Frobenius inner product \cite{Belitskii02,Elphick87}. 
For each $p,q \in \Pnd$, define
\begin{equation} \label{eq:inpro}
	\langle p , q \rangle  \equiv p(\partial_\xi) \cdot q(\xi)|_{\xi=0} \,,
\end{equation}
where ``$\cdot$" denotes the Euclidean scalar product and $p(\partial_\xi)$ is the polynomial $p$ with each occurrence of $\xi_i$ replaced by the derivative $\frac{\partial}{\partial \xi_i}$.
For example, for two vector monomials \eqref{eq:monomialBasis}, 
\[
	\langle p_{i,m}, p_{j,\tilde m} \rangle = m! \delta_{m,\tilde m} \delta_{i,j}\;, 
\]
where $m! \equiv m_1! m_2! \ldots m_n!$. It is easy to see that $\langle p, q\rangle = \langle q, p \rangle$, and $\langle p, p \rangle > 0$ when $p \neq 0$. For
linear vector fields, $d=1$, this inner product reduces to
\[
    \langle A\xi , B \xi \rangle = \sum_{i,j=1}^n A_{ij} B_{ij} \;,
\]
the inner product that defines the Frobenius norm for matrices. 

In this norm, the two subsets of basis vectors
for $\Vnd$ are orthogonal:

\begin{lem}
The basis vectors $p_{i,\check{m}}$, $v_{j,\bar{m}}$ of $\Vnd$ are orthogonal for any
$i,j \in [1,n]$ and $m,\bar{m} \in \N^d$.
\end{lem}

\begin{proof}
This follows because $p_{i,\check{m}}$ is missing $\xi_i$ in the $i^{th}$ component, but $v_{j,\bar{m}}$ always has $\xi_i$ in component $i$ if that component is non-vanishing.
\end{proof}
 
We define one additional subspace of $\Fnd$:
\begin{equation}\label{eq:UndDefine}
	\Und = \{ \theta(\xi) \xi: \theta(\xi) \in \Pnd \} \;.
\end{equation}
A basis for $U^n_d$ is given by the set vector fields
\[
    u_{m} = \xi^m \sum_{i=1}^n  \xi_i \hat{e}_i = \xi^m \xi \;, \quad |m| = d-1 \;.
\]
For example,
\begin{equation}\label{eq:P32basisU}
 U^3_2 = \mbox{span} \left\{ 
            \begin{pmatrix} x^2 \\ xy \\ xz \end{pmatrix} \;,
            \begin{pmatrix} xy \\ y^2 \\ yz \end{pmatrix} \;,
            \begin{pmatrix} xz \\ yz \\ z^2 \end{pmatrix}
             \right \} \;.
\end{equation}
Note that each basis vector in $\Und$ has non-vanishing divergence $\nabla \cdot u_m = (n + d-1)\xi^m $; indeed, this space is the complement of the divergence free space.

\begin{lem} \label{lem:block}
For the inner product \eqref{eq:inpro}, the spaces $\Und$ and $\Vnd$ are orthogonal complements in $\Fnd$.
\end{lem}
\begin{proof}
For any $v \in \Vnd$ and $u = \theta(\xi) \xi \in \Und$ we have
\[ 
	\langle \theta(\xi) \xi, v \rangle = \langle \theta, \nabla \cdot v\rangle = 0 \;.
\]
Thus $\Und$ and $\Vnd$ are orthogonal.
That they are complementary spaces simply follows from the observation that their
dimensions add to that of $\Fnd$. By \eqref{eq:UndDefine}, the dimension of $\Und$ is that same as that
of degree $d-1$ polynomials, 
\[
	\dim \Und = \dim P^n_{d-1} \;.
\]
Therefore, \eqref{eq:dimenVnd} gives $\dim \Fnd = \dim \Vnd + \dim \Und$.
\end{proof}

\begin{rem}
For $n=3$, the orthogonal decomposition $F^3_d= V^3_d \oplus U^3_d$ seems  reminiscent of the Helmholtz decomposition of vector fields, $h = \nabla \times \psi + \nabla \phi$. However, it is different---in particular the curl of a vector $\theta \xi$ in $U^3_d$ does not vanish in general.
The difference is that for the Helmholtz decomposition, the scalar product between
vector fields is defined by the integral of the Euclidean scalar product (in $\R^3$).
Nevertheless, the dimension of the space of gradient vector fields that are not
harmonic vector fields is $\dim P_{d+1}^3 - (2(d+1) + 1) = \dim P_{d-1}^3$, the same dimension as $U^3_d$.
\end{rem}

\section{Normal form for vector fields}\label{sec:VectorField}

In this section we will use the inner product style to compute the complement to $\rng{\Lvf}$.\footnote
{
	In our case this gives the same result as the $\sl(2)$ style
}
This will give the formal normal form for a vector field with a triple-zero eigenvalue.
The results here reproduce those of \cite{Elphick87}. We will use an intermediate form
of this result in the next section to project onto the divergence-free case and complete the proof of
\Th{thm:VF}.

The main advantage of the inner product \eqref{eq:inpro} is that it
allows a simple construction of the adjoint of $\Lvf_N= \ad_N$ on the space of polynomial vector fields \cite{Belitskii02,Elphick87}. Along the way,
we prove the analogous result for the operator $\Ad_J$ that will be used
for the map case.

\begin{lem} \label{lem:adjoint}
	Using the inner product \eqref{eq:inpro}, $(\Ad_J)^* = \Ad_{J^*}$ and $(\ad_N)^* = \ad_{N^*}$.
\end{lem}

\begin{proof}
From the definition \eqref{eq:inpro}
\[
   \langle p , \Ad_J q \rangle = p(\partial_\xi) \cdot J \; q(J^{-1}\xi)|_{\xi=0} \;.
\]
Change coordinates to $\eta = J^{-1}\xi$, noting that $\partial_\xi = J^{-1*} \partial_\eta$ to obtain
\begin{equation}\label{eq:Adjoint}
   \langle p , \Ad_J q \rangle = (J^* p(J^{-1*}\partial_\eta)) \cdot q(\eta)|_{\eta=0}
   = q(\partial_\eta) \cdot (J^* p(J^{-1*} \eta))|_{\eta=0} = \langle q, \Ad_{J^*} p \rangle \;.
\end{equation}
The analogous statement for $\ad_N$ follows from $\Ad_{e^{tN}} = e^{t \ad_{N}}$ and $(e^{tN})^* = e^{tN^*}$. Setting $J = e^{tN}$ in \eqref{eq:Adjoint} and 
differentiating with respect to $t$ at $t=0$ gives the result.
\end{proof}

This lemma implies that  $\coker{\ad_N} = \ker \ad_{N^*}$ is the orthogonal complement to the range of $\Lvf_N$. Using \eqref{eq:adDefine}, the cokernel of $\ad_N$ is therefore determined by the solutions of
\begin{equation}\label{eq:adStar}
	\ad_{N^*} h = \cD_{N^*} h - N^* h  = 0 \;,
\end{equation}
where we introduce the linear operator $\cD_N \equiv N \xi \cdot \nabla$.

Here we solve \eqref{eq:adStar} for the $n$-dimensional generalization of \eqref{eq:3DN}
\begin{equation}\label{eq:nDN}
	N = \begin{pmatrix} 0 & 1 & 0 & 0 &\ldots \\
					   0 & 0 & 1 & 0 &\ldots \\
					   0 & 0 & 0 & 1 &  0 \\
					   \vdots &\vdots &\vdots & 0 & \ddots \\
		 \end{pmatrix} \;,
\end{equation}
that is, $N_{i,i+1} = 1$, and $N_{ij} = 0$ otherwise.
For the adjoint of $N$, the linear operator $\cD$ becomes
\[
	\cD_{N^*}  = \sum_{i=1}^{n-1}\xi^i \frac{\partial}{\partial \xi^{i+1}}\;,
\]
and \eqref{eq:adStar} is equivalent to the system of linear PDEs
\begin{equation} \label{eq:sysPDE}
	\cD_{N^*} h_1 = 0\;, \quad 
	\cD_{N^*} h_j = H_{j-1} \;,\; j = 2,3,\ldots n\;.
\end{equation}
This system is easily solved by the method of characteristics. The solution
of the characteristic equations $ \dot{\xi} = N^* \xi$ and $\dot{h} = N^* h$ is formally easy:
\begin{align}\label{eq:charsol}
	\xi(t) &= e^{tN^*} \xi^0 \;,\nonumber\\
	h(t) &= e^{tN^*} h^0 \;.
\end{align}
A solution of the PDE is obtained by inverting the equations for $\xi(t)$ to solve
for the invariants $\xi^0 = e^{-tN^*}\xi$. However, we must use one of the equations to eliminate time. The easiest way to do this is to use the equation for the second variable since it is linear in $t$. Denote the first three coordinates by $(x,y,z)$ so that  $\xi = \left(x,y,z,\ldots,\xi_n \right)$. Then \eqref{eq:charsol} gives $y(t) = y^0 + t x^0$. When $x(t) = x^0 \neq 0$ the characteristics all pass through the surface $y^0 = 0$. So up to this singularity we can choose this surface to define the initial conditions. To do this, set $t = y/x$, to obtain the invariants
\begin{equation}\label{eq:invariants}
 \xi^0(x,y,\ldots \xi_n) = e^{-\frac{y}{x} N^*} \xi \;.
\end{equation}
These invariants are rational functions with denominator $x^{k-1}$. The first invariant is simply $x$ itself, and the solution of the PDE will depend on arbitrary functions of the invariants. Thus we can clear the denominators in the remaining equations and define $n-1$ polynomial invariants
\begin{align}\label{eq:3Dinvariants}
	\alpha  &= \xi^0_1 = x \;, \nonumber\\
	\beta &= 2x\xi^0_3 = 2zx - y^2 \;, \nonumber\\
	\gamma &= 3x^2 \xi^0_4 =  3x^2w - 3xyz + y^3 \;,
\end{align}
and so forth. 
Therefore, a formal solution to \eqref{eq:sysPDE} is
\begin{equation} \label{eq:Hgen}
   h(\xi) = e^{y/x N^*}  \phi(x,\beta,\gamma,\ldots) \;,
\end{equation}
for an arbitrary vector valued function $\phi$ of the $n-1$ invariants.
For example, for $n=2$ the solution is
\begin{equation}\label{eq:H2D}
    h = \begin{pmatrix}
		\phi_1(x) \\
		\frac{y}{x}\phi_1(x)  + \phi_2(x) \\
		\end{pmatrix} \;,
\end{equation}
and when $n=3$ we obtain
\begin{equation}\label{eq:H3D}
    h = \begin{pmatrix}
		\phi_1(x, \beta) \\
		\frac{y}{x}\phi_1(x,\beta)  + \phi_2(x, \beta) \\
		\frac{ y^2}{2x^2} \phi_1(x,\beta) +
			 \frac{y}{x} \phi_2(x,\beta)  + \phi_3(x, \beta)
    \end{pmatrix} \;.
\end{equation}

The desired solution for $h$ is in $\Fnd$. It is, however, a nontrivial problem to obtain the most general polynomial solution. Even if we assume that the functions $\phi_i$ are polynomials, the solution \eqref{eq:Hgen} is generally rational because it contains powers of $t=y/x$. For $n=2$ it is clear that $\phi_1(x)$ must be $x$ times a polynomial and the solution is simply
\begin{equation}\label{eq:elphick2D}
   h = \begin{pmatrix} 
			x \varphi_1 \\ y \varphi_1 + \varphi_2 \;,
		\end{pmatrix}
\end{equation}
where $\varphi_i$ are polynomials that depend upon $x$ only. 

For $n=3$ one might a first think that $\phi_1 = x^2 \varphi_1$ is required,
since it appears with denominator $x^2$ in the third component of \eqref{eq:H3D}.
However, this is not the most general polynomial solution, since we can replace $y^2$ by $2zx -\beta$ to eliminate one power of $x$ in the denominator and then remove the second by setting $\phi_1 = x \varphi_1$ and 
$\phi_3 = \frac{\beta}{2x} \varphi_1 + \varphi_3$. This changes
the third term to $ z \varphi_1 + \frac{y}{x} \phi_2
+ \varphi_3$. Finally setting $\phi_2 = x \varphi_2$, we obtain the 
polynomial solution for $n=3$
\begin{equation}\label{eq:elphick3D}
   h = \begin{pmatrix} 
	x \varphi_1 \\ y \varphi_1 + x \varphi_2 \\ z \varphi_1 + y \varphi_2 + \varphi_3
   \end{pmatrix}\;,
\end{equation}
where $\varphi_i = \varphi_i(x, \beta)$ are polynomials of appropriate degree.
This solution was shown to be the general polynomial solution in \cite{Elphick87}.

With these forms it is easy to obtain the dimensions of the kernel
of $ad_{N^*}$ for two and three dimensions, giving the entries in the $n=2$ and $3$ rows of the leftmost pane of \Tbl{tbl:dimtab}.

\begin{lem}\label{lem:dimkerad}
The dimension of $\ker \ad_{N^*}$ in $P^2_d$ is $2$ and in $P^3_d$ is  $\lceil 3d/2+1 \rceil$.
\end{lem}
\begin{proof}
This is a simple counting argument, based on the polynomial forms.
For $n=2$ at degree-$d$, $h$ contains two monomials, $\varphi_1 = a x^{d-1}$ and $\varphi_2 = b x^d$
so the dimension is always two.

For $n=3$, the polynomials $\varphi_i$ in \eqref{eq:elphick3D} whose arguments are of degree one and two, respectively, must be chosen appropriately. The even and odd cases can be treated separately: a degree-$2m$ polynomial has the form
\begin{equation}\label{eq:evenPoly}
	\sum_{l=0}^{m} a_l x^{2l}\beta^{m-l}  
		\in P^3_{2m}\;,
\end{equation}
and a degree $2m+1$ polynomial has the form
\begin{equation}\label{eq:oddPoly}
	\sum_{l=0}^{m} b_l x^{2l+1}\beta^{m-l} 
		\in P^3_{2m+1} \;.
\end{equation}
Each of these sums has $m+1$ arbitrary coefficients, so the degree-$d$ case has $\lceil \frac{d+1}{2}\rceil$ coefficients.

To apply this to the form \eqref{eq:elphick3D}, note that when $h$ is degree $d$, then $\varphi_1$ and $\varphi_2$ have degree $d-1$. Thus each has $\lceil \frac{d}{2} \rceil$ coefficients. The function $\varphi_3$ is degree $d$, so it has $\lceil \frac{d+1}{2} \rceil$ coefficients. Thus total number of coefficients is
\[
	2\lceil \tfrac{d}{2} \rceil + \lceil \tfrac{d+1}{2} \rceil = \lceil \tfrac{3d}{2}+1 \rceil \;.
\]
Since the each monomial in each function represents an independent vector in $h$, this is the same as the dimension.


\end{proof}

\begin{rem}
The construction of a general polynomial solution to \eqref{eq:sysPDE} when $n>3$ is complicated by the fact that there are polynomial combinations of the invariants that have $\alpha=x$ as a factor. For example the combination $\beta^3 + \gamma^2 = \alpha^2\delta$ where $\delta$ is a quartic polynomial in $(x,y,z,w)$. Consequently, the power of $x$ in the denominator of terms involving these combinations is different than its nominal value. It can be shown that is the only nontrivial relation in the case $n=4$, and this leads to the polynomial form for this case \cite{Murdock03}. We are not aware that $n=5$ has been explicitly worked out. Thus the dimension of these subspaces is harder to compute when $n > 3$ (see \Tbl{tbl:dimtab}, for $n \ge 4$).
\end{rem}

\begin{table} 
\caption{Dimensions of the various (co)kernels for the homological operator $ad_N$ with $N$ in \eqref{eq:nDN} in dimension $n$ at degree $d$. \Th{thm:VF} for $n=2,3$ implies that the first two rows of the middle and right tables coincide for any $d$.}
\label{tbl:dimtab}
\[
\begin{array}{l|lllll}
\multicolumn{6}{c}{\dim (\ker  \ad_N)}\\
n \setminus d 	& 2 & 3 & 4 & 5 & 6\\ \hline
2      			& 2  & 2 & 2 & 2 & 2 \\
3      			& 4  & 6 & 7 & 9 & 10\\
4      			& 7  & 12 & 17 & 24 & 31 \\
5      			& 11 & 21 & 36\\
6      			& 15 \\
\end{array}
\quad
\begin{array}{l|lllll}
\multicolumn{6}{c}{\dim (\ker \ad_N \cap \Vnd)}\\
n \setminus d & 2 & 3 & 4 & 5 & 6\\ \hline
2      & 1  & 1 & 1 & 1 & 1\\
3      & 3  & 4 & 5 & 6 & 7\\
4      & 6  & 10 & 14 & 19 & 25 \\
5      & 10 & 18 & 31\\
6      & 14 \\
\end{array}
\quad
\begin{array}{l|lllll}
\multicolumn{6}{c}{\dim P^{n-1}_d }\\
n \setminus d & 2 & 3 & 4 & 5 & 6\\ \hline
2      & 1  & 1 & 1 & 1 & 1\\
3      & 3  & 4 & 5 & 6 & 7\\
4      & 6  & 10 & 15 & 21 & 28 \\
5      & 10 & 20 & 35 \\
6      & 15 \\
\end{array}
\]
\end{table}

\section{Divergence free vector fields}\label{sec:DivergenceFree}

In this section we complete the proof of \Th{thm:VF}. The remaining task is
to compute a complement to the range of $\ad_N$ for divergence-free vector fields. We
start by showing that for any matrix $N$, the splitting $\Fnd = \Vnd \oplus \Und$ block diagonalizes $\ad_N$, that
is
\begin{equation}\label{eq:blockDiag}
	\ad_N \Vnd \subset \Vnd\;, \mbox{ and } \ad_N \Und \subset \Und \;.
\end{equation}
Moreover, since $(\ad_N)^* = \ad_{N^*}$ this also applies to $\ad_{N^*}$.

This block diagonalization allows us to compute $\coker \ad_N$ for divergence-free vector fields by simply projecting the full $\coker$ onto $\Vnd$. This projection can be accomplished by solving an ODE for one of the components of $h$. However, the result is not the simplified normal form of \Th{thm:VF}. The last step is to show that the simplified form is also a complement to $\rng{\ad_N}$ by showing that its projection onto $\coker \ad_N$ is of full rank. 

We begin by verifying \eqref{eq:blockDiag} in the next two lemmas.

\begin{lem} \label{lem:Vinva}
The volume preserving subspace $\Vnd$ is an invariant subspace of $\ad_N$ for any matrix $N$; in other words, $\ad_N \Vnd \subset \Vnd$.
\end{lem}

\begin{proof}
Our goal is to show that if $\nabla \cdot v = 0$ then $\nabla \cdot (\ad_N v) = 0$. 
Writing the latter in components gives
\[
	\partial_i (\ad_N v)_i
	= \partial_i ( N_{km} \xi_m \partial_k v_i - N_{ik} v_{k}) 
	= [ Dv, N ]_{ii} +  (N \xi \cdot \nabla) \partial_i v_i  \;,
\]
where $[\;,\;]$ is the matrix commutator and we use the summation convention.
The first term is simply the trace of the commutator of $Dv$ and $N$, but
$\tr [A,B] = 0$ for any two matrices. The last term is $(N\xi \cdot \nabla) (\nabla \cdot v)$, which
vanishes since $v \in \Vnd$.
\end{proof}

\begin{rem}
Notice that it was \emph{not} necessary to assume $\tr N = 0$ in this Lemma because
the trace of the commutator of any two matrices vanishes.
For other Lie algebras, e.g.\  $\sp$ or $\so$, it is necessary to assume that $N$ is in the algebra to get a block diagonalization since otherwise the 
commutator $[Dv, N]$ is not in the algebra.%
%
\footnote{
Preserving symmetry in the normal form is different.
There a matrix group $G$ exists so that $[G, N] = 0$. 
This is replacing the condition on the differential of $K$ and on $N$.
When $G$ is unitary this implies $[G, N^*] = 0$. 
However, there is no associated invariant subspace, 
instead we find $[\Ad_G, \ad_N] = 0$  and $[\Ad_G, \ad_{N^*}] = 0$, 
which means that the cokernel is invariant under the action of $G$.
This implies that the normal form has the same symmetry group.
However, this does not mean that there is a subspace invariant under $\ad_N$.
}
\end{rem}

This lemma implies that the operator $\ad_N$ acting on a vector in components $v+u$ is block upper-triangular. To show that it is block diagonal, we must also show that $\ad_N$ leaves the orthogonal complement $\Und$ invariant.

\begin{lem} \label{lem:Uinva}
The non-volume-preserving subspace $\Und$ is an invariant subspace of $\ad_N$ for any matrix $N$. Thus $\ad_N \Und \subset \Und$.
\end{lem}

\begin{proof}
From \eqref{eq:UndDefine}, the general element of $\Und$ has the form $\theta(\xi) \xi$ where $\theta \in \Pnd$ is
a scalar polynomial. Such a vector remains in $\Und$ since
\[
    \ad_N \theta \xi = D(\theta \xi) N \xi  - N \theta \xi = 
    		(\nabla \theta \cdot N \xi )  \xi\;,
\]
is again of the form of scalar, degree-$d$ polynomial times $\xi$.\end{proof}

Thus $\coker \ad_N$ can be split into two subspaces
\[
	\coker \ad_N = (\coker \ad_N \cap \Und) \oplus (\coker \ad_N \cap \Vnd) \;,
\]
that are invariant under $\ad_N$. Thus, for a divergence-free vector field \eqref{eq:original}, the solution $h$ to the homological equation \eqref{eq:homoEq} can be taken to be itself divergence free, and the normal form can be selected to be in $\coker \ad_N \cap \Vnd$.

We will only construct these subspaces for the two and three-dimensional cases.

The two-dimensional case is simple.
The general solution of the homological equation in this case was given in \eqref{eq:H2D}. Imposing $\nabla \cdot h = 0$ implies that $\phi_1(x) = 0$, so that $h = \hat{e}_2 \phi_2(x)$. This implies that $\dim \coker \ad_N \cap V^2_d = 1$, as shown in Table~\ref{tbl:dimtab}. Note that the resulting vector field already has the ``simplified" form of \Th{thm:VF}. In this case, we could also have imposed the volume-preserving condition on the polynomial solution \eqref{eq:elphick2D} and obtained the same result. Neither of these two statements are true for the three-dimensional case

\begin{lem} \label{lem:cokerV}
The intersection $(\coker \ad_N)\cap V^3_d$ is the set of vector fields of the form
\begin{equation}\label{eq:cokerV}Â
   h = \begin{pmatrix}
           x^2 \psi_1 \\
           x y \psi_1 + x \psi_2 \\
           \frac12 y^2 \psi_1 + y \psi_2 + \psi_3
   \end{pmatrix}\;,
\end{equation}
where $\psi_i = \psi_i(x, \beta) \in P^3_d$ and $\psi_3$ satisfies
\begin{equation}\label{eq:divFreePsi}
   -2 \partial_\beta \psi_3 =  x \partial_x \psi_1 + 3 \psi_1 + \beta \partial_\beta \psi_1 \,.
\end{equation}
The dimension of the cokernel of $\ad_N$ restricted to $V^3_d$ is $d+1$ as
in \Tbl{tbl:dimtab}.
\end{lem}

\begin{proof}
An explicit computation of the divergence of (the not necessarily polynomial solution) \eqref{eq:H3D} gives 
\[
    - 2\partial_\beta \phi_3 = \frac{1}{x} \partial_x \phi_1 + \frac{1}{x^2}( \phi_1 + \beta \partial_\beta \phi_1) \;.
\]
For any given $\phi_1$ and $\phi_2$, this ODE for the dependence $\phi_3$ on $\beta$ can be integrated with respect to $\beta$. For $\phi_3$ to be polynomial we require $\phi_1 = x^2 \psi_1$.
Similarly we need to set $\phi_2 = x \psi_2$ in order to make the $2^{nd}$ component 
of $h$ polynomial.

The polynomial $\psi_1$ has degree $d-2$, so by \eqref{eq:evenPoly} and $\eqref{eq:oddPoly}$ it has $\lceil \frac{d-1}{2} \rceil$ coefficients. Similarly $\psi_2$ is of degree $d-1$ and has $\lceil \frac{d}{2} \rceil$ coefficients.  For given $\psi_1$ and $\psi_2$,  $\psi_3$ is determined up to an arbitrary function of $x$, which introduces a single term $c x^d$ at fixed degree $d$.
Thus the total number of coefficients is
\[
	\lceil \frac{d-1}{2} \rceil + \lceil \frac{d}{2} \rceil + 1 = d+1 \;.
\] 
\end{proof}

For completeness, we also can obtain a representation for the
non-volume-preserving block.

\begin{lem} \label{lem:cokerU}
The intersection $(\coker \ad_N) \cap U^3_d$ is the set of vector fields
$\{ \theta \xi : \theta(x,\beta) \in P^3_{d-1}, \beta = 2zx-y^2 \}$. This space has dimension  $\lceil \frac{d}{2} \rceil$.
\end{lem}
\begin{proof}
From \Lem{lem:Uinva} we see that we need to solve the equation 
\[
    \ad_{N^*} \theta \xi = ( \nabla \theta \cdot N^* \xi) = \cD_{N^*} \theta  = 0 \,;
\]
which is the same as the first PDE in the system \eqref{eq:sysPDE}.
The method of characteristics, as in \Sec{sec:VectorField} implies that the general solution is an arbitrary function of the invariants \eqref{eq:invariants} as before. When $n=3$ the invariants $x$ and $\beta$ have no combinations that have $x^k$ as a factor, so the general polynomial solution is
\[
	u = \theta(x,\beta) \xi\;,
\]
for a polynomial function $\theta$. Since $\theta$ has degree $d-1$,  \eqref{eq:evenPoly} and \eqref{eq:oddPoly}, imply that it has $\lceil \tfrac{d}{2} \rceil$ coefficients.
\end{proof}

Lemma~\ref{lem:cokerV} shows that that the dimension of $\ker \ad_N^* \cap V^3_d$ 
is the same as that of the space of a single polynomial in two variables (given in the right pane of \Tbl{tbl:dimtab}).  This makes it plausible that the divergence-free vector fields of the form $h = (0,0,p(x,y))^T$ form a possible complement of the range of $\ad_N$. This is proved in the next lemma. 

The simplified normal form was introduced by Murdock for the general case \cite{Murdock03}. He argued that a simplified complement to the range of $ad_N$ has the form of \eqref{eq:elphick3D} but with $h_1 = h_2 = 0$. We show next that a similar projection can be done in the divergence-free case. 

\begin{lem} \label{lem:newcoker}
The set $\{ \hat{e}_3 p(x,y): p \in P^2_d \}$ is a complement to $\rng \ad_N \cap V^3_d$. 
\end{lem}
\begin{proof}
We need to demonstrate that the orthogonal projection of the new set to the volume-preserving subspace $\coker \ad_N \cap \Vnd$ is 1-to-1. In other words, the matrix of inner products of  bases for the two spaces is nonsingular.

Consider bases for the two subspaces. 
A basis for the simplified complement are the $d+1$ vector monomials $\{ \hat{e}_3 x^{k}y^{d-k} \}$ where $k = 0,1,
\ldots d$.

The basis for $\coker \ad_N \cap V^3_d$ given in \eqref{eq:cokerV}
can be constructed by using monomials for the functions $\psi_1$ and $\psi_2$.
For example when $d=2m$ is even, the monomial basis elements correspond to $\psi_1 = x^{2j-2} \beta^{m-j}$ and $\psi_2 = x^{2j-1}\beta^{m-j}$, for $j = 1,2,\ldots m$. The final basis element corresponds to the free function obtained from integrating \eqref{eq:divFreePsi}, $\psi_3 = x^{2m}$. The $d+1$ basis vectors are then
\[
	h_{1,j} = \begin{pmatrix} x^{2j} \beta^{m-j} \\ x^{2j-1}y \beta^{m-j} \\ 
			\frac12 \left(y^2 - \frac{m+j+1}{m-j+1}\beta \right) x^{2j-2}\beta^{m-j} \end{pmatrix} \;, \quad
	h_{2,j} = \begin{pmatrix} 0 \\ x^{2j} \beta^{m-j} \\ x^{2j-1}y \beta^{m-j} \end{pmatrix} \;, \quad
	h_3 = \begin{pmatrix} 0 \\ 0 \\ x^{2m} \end{pmatrix} \;.
\]
The inner product of a general basis vector with $h_{1,j}$ gives
\begin{align*}
	\langle \hat{e}_3 x^{k}y^{2m-k}, h_{1,j} \rangle &= 
	\langle x^{k}y^{2m-k}, 
		\frac12 (y^2 - \frac{m+j+1}{m-j+1}\beta) x^{2j-2}\beta^{m-j} \rangle \\
	&=	-\frac{m+1}{m-j+1} \langle x^{k}y^{2m-k},  x^{2j-2}(-y^2)^{m-j+1} \rangle \;. 
\end{align*}
This is clearly nonzero only for $k=2j-2$, that is when $k = 0,2,\ldots, d-2$.
The second set of functions $h_{2,j}$ have inner products with the simplified
basis given by
\begin{align*}
	\langle \hat{e}_3 x^{k}y^{2m-k}, h_{2,j} \rangle &= 
	\langle x^{k}y^{2m-k} ,  x^{2j-1}\beta^{m-j} \rangle \\
	&=	 \langle x^{k}y^{2m-k} ,  x^{2j-1}(-y^2)^{m-j} \rangle \;.
\end{align*}
which is nonzero only for $k = 2j-1$, or $k = 1,3,\ldots d-1$.
The final inner product $\langle \hat{e}_3 x^k y^{2m-k}, h_3 \rangle$ is nonzero when $k =2m$. 
Hence up to a permutation the matrix of scalar products between the basis vectors is 
diagonal and thus has full rank.
Thus the orthogonal projection of the set $\hat{e}_3 p(x,y)$ to $\coker \ad_N \cap V^3_d$ is 1-to-1. 
A similar calculation pertains for the case that $d = 2m+1$ is odd; in this case the monomials in $\psi_1$ cover the odd powers of $x$ and the monomials in $\psi_2$ cover the even powers.


We have now proved \Th{thm:VF}.
\end{proof}

\begin{rem}
Note that while Murdock's simplified normal form (obtained by setting $h_1 = h_2 = 0$ in \eqref{eq:elphick3D}) 
\begin{equation}\label{eq:Murdock}
    h(x,y,z) = \hat{e}_3 (z \varphi_1(x,\beta) + y \varphi_2(x,\beta) + \varphi_3(x,\beta))\;,
\end{equation}
does form a complement to $\rng \ad_N$ for the general, non-volume preserving case, we cannot simply project 
the simplified normal form onto the volume-preserving subspace. Indeed, 
the only degree-$d$, volume-preserving elements for the simplified normal form are
\[
	h(x,y,z) 
			 = \hat{e}_3 (a y^2 x^{d-1}+ byx^{d-1} + cx^{d}) \;,
\]
giving a dimension of three, which is too small (except for $d=2$).  The point is that the transformation from inner product form to simplified normal form is equivalent to changing the scalar product. The new scalar product is defined by declaring the range of $\ad_N$ and the proposed complement to be orthogonal. Assuming that the dimensions are right this is always possible. However, the new scalar product need not respect the divergence free conditions. More precisely, when the adjoint of $\ad_N$ is defined with respect to this scalar product, then $\Vnd$ and $\Und$ need not be invariant subspaces. 
\end{rem}

\begin{rem}
We have shown that $\ad_N$ can be block-diagonalized into 
the divergence free subspace and its orthogonal complement. 
The basis given in \Sec{sec:gen} achieves this block-diagonalization
by \Lem{lem:block}. Thus it is interesting to compare the different representations
of the complement of the kernel. We have just proved that another ``simplified"  complement is
\[
     \tilde h  = \hat{e}_3 p(x,y) + \theta (x, \beta)\xi \;,
\]
which is different from \eqref{eq:Murdock}.
Nevertheless, the projection argument works in this case as well.
\end{rem}

\section{Volume preserving maps}\label{sec:VPMaps}

The normal form that we have just obtained for divergence-free vector fields can essentially be transferred to volume-preserving mappings by exponentiating $\ad_N$ to obtain the homological operator for \eqref{eq:homoMap}. In this section we adapt the results of Bridges and Cushman \cite{BridgesCushman93} for Hamiltonian vector fields to divergence-free vector fields. The main difference is that in the volume-preserving case there is no scalar generating function.

There are three basic steps to transferring \Th{thm:VF} to the mapping result \Th{thm:MAP}. First we
note that the mapping homological operator $\Lmap$ has the same adjoint properties
as the flow operator $\Lvf$ with respect to the inner product \eqref{eq:inpro}. Then we 
use the fact that $\Ad$ is the exponential of $\ad$ to relate the fundamental spaces of these operators.
Finally we show that it is possible to choose coordinates in which the linear form of the map is the exponential of the nilpotent block $N$ for the flow.

As for the vector-field case, we will use the inner product style. Recall that 
\Lem{lem:adjoint} implies that $(\Ad_J)^* = \Ad_{J^*}$ using the
inner product \eqref{eq:inpro}. Thus
the mapping homological operator \eqref{eq:homoMap} satisfies
\[
	(\Lmap_J)^* = (\Ad_{J^{-1}} - \Id)^* = \Ad_{J^{-1*}} - \Id = \Lmap_{J^*} \;.
\]
Consequently, the construction of the normal form for the mapping \eqref{eq:mapOriginal} reduces to finding a representation for $\ker\Lmap_{J^*}$. Recall that $\Ad_{e^N} = e^{\ad_N}$, thus there is a relation between the homological operators:
\[
       \Lmap_J = \Ad_{J^{-1}} - \Id = \exp ( \ad_{-\log J} ) - \Id\;,
\]
providing $J$ has a (real) logarithm. This is true for \eqref{eq:mapNF} since none of the eigenvalues of $J$ are negative. We will thus consider $\ad_{-\log J}$  as the appropriate homological operator instead of the more difficult homological operator for maps. 

To use the relation between $\Ad$ and $\ad$, we need to relate their fundamental spaces. The basic lemma relates the spaces for any nilpotent operator. This crucial step was inspired by \cite{BridgesCushman93}.

\begin{lem}\label{lem:nilpotentrange}
	If $N$ is a nilpotent matrix then $M = \exp{N}-\Id$ and $N$ have the same kernel and range.
\end{lem}
\begin{proof}
	Note that $M = N(\Id+\frac12N + \ldots \frac1{m!}N^{m-1}) = NB = BN$.
$B$ is a polynomial in $N$ because $N$ is nilpotent.
We claim that $B$ is nonsingular. Indeed when $N$ is in Jordan form, then $B$ is upper triangular and has a diagonal of all ones.  Now, if $w \in \rng{M}$ then there is a $v$ such that $w = Mv = NBv$, so $w \in \rng{N}$. Moreover, if $w \in \rng{N}$ then $w = Nv = (NB)B^{-1}v = M B^{-1}v$ so $w \in \rng{M}$. Thus $\rng{M} = \rng{N}$. 

Similarly if $z \in \ker{M}$ then $0 = B^{-1}Mz = Nz$ and if $z \in \ker{N}$ then
$0 = BNz =Mz$. So $\ker{M} = \ker{N}$.
\end{proof}

In order to apply the previous lemma to $\ad_N$ we need the following well-known fact about adjoint representation of nilpotent matrices.
\begin{lem} \label{lem:adNisnil}
	If $N$ is nilpotent, then $ad_N$ is nilpotent.
\end{lem}
\begin{proof}
This follows from the fact that there exists a triad $N,M,K$ that 
form a representation for $\sl(2)$, i.e., that $[N,M] = K$, $[K,N]= 2N$, and $[K,M]=-2M$ and for any such representation, the matrices $N$ and $M$ are nilpotent and $K$ is semisimple. Finally the operators $ad_N$, $ad_M$ and $ad_K$ also form such a representation. 
This proof is along the lines of Murdock \cite{Murdock03} Thm 2.5.2, but it is 
a general theorem, see e.g.~\cite{Serre87}.
\end{proof}

The last step fixes the problem that \Th{thm:VF} is stated only for nilpotent Jordan blocks $N$,
but $\log J = \log (\Id+ N)$ is not of this form. The remedy is to first do a linear change of coordinates
that puts the volume-preserving map into a form where its linear part is $\exp N$
instead of $\Id + N$. Then \Th{thm:MAP} connects to \Th{thm:VF}. 
The transformation must be chosen so that it does not destroy the simplified form 
of the map, in which all the nonlinearity is concentrated in the last component.

\begin{lem}\label{lem:expConj}
	Suppose $N$ is a nilpotent matrix, then there is an invertible matrix $T$
	such that $T \exp{N} T^{-1} = I+N$.
	Moreover the conjugacy can be chosen such that $T$ leaves $\hat{e}_n$ invariant.
\end{lem}
\begin{proof}
While this lemma can be proved by induction for arbitrary dimensions, since we will
apply it only for $n=2$ and $n=3$, we simply give the matrices for these cases. 
For $n=2$ there is nothing to do since $N_2^2 = 0$, so that $\exp N_2 = \Id + N_2$.
For the $3 \times 3$ case we can see that
\[
    \exp{N_3} = \Id + N_3 + \frac12 N_3^2 = \begin{pmatrix} 1 &1 & \frac12 \\
    							 0 & 1 & 1 \\
							 0 & 0 & 1
			  \end{pmatrix}\;.
\]
We have $T_3 \exp{N_3} = (\Id+N_3)T_3$ using 
\[
	T_3 = \begin{pmatrix} 	1 &-\frac12 & 0 \\
						0 & 1 & 0 \\
						0 & 0 & 1 
			  \end{pmatrix}\;.
\]

\end{proof}


Thus, using the new coordinate system, we can apply \Th{thm:VF} to the map case, giving
the promised normal form \eqref{eq:mapNF}.

To finish the proof of \Th{thm:MAP} we only need to verify the claims about the
properties of the normal form. Firstly every truncation of the map is {\em exactly} volume
preserving. The determinant of the Jacobian of $g(\xi) = J(\xi + \hat{e}_3p(x,y))$ is the product of 
$\det J = 1$ and $\det (\Id + D \hat{e}_3 p(x,y)) = 1$, since $D\hat{e}_3 p(x,y)$ is lower triangular.
Another way of viewing this result is to say that time-one map of the vector field
$\hat{e}_3 p(x,y)$ is the map $\xi + \hat{e}_3 p(x,y)$.
Secondly the inverse of the map can be explicitly computed. 
Let $\xi' = g(\xi)$. Then
\[
     \xi = g^{-1}(\xi') = J^{-1} \xi' - \hat{e}_3 Q(x, y) = J^{-1} \xi' - \hat{e}_3(x'-y'+z', y'-z') \;.
\]
This completes the proof of \Th{thm:MAP}.

\section{Unfolding}\label{sec:unfolding}

The unfolding of the triple eigenvalue collisions treated above is obtained
by considering a family depending on $\epsilon$ and additional unfolding
parameters $\mu_i$. The parameter $\epsilon$ will be used to unfold the 
saddle-node bifurcation, while the parameters $\mu_i$ will ensure that 
an arbitrary spectrum of the linearization can obtained near the bifurcation 
$\epsilon = \mu_i = 0$. The original map is now considered to depend upon 
$\epsilon$, and is expanded in a power series in $(\epsilon,\mu)$. 
The zero-order terms have been treated in the previous section. 
Normalization now proceeds order by order in $(\epsilon,\mu)$. 
The main observation (see, e.g.~\cite{Elphick87}) is that at every order in the
parameters the same homological operator is obtained. Thus all the results obtained above
hold at each order, and hence instead of the polynomial $p(x,y)$ with constant 
coefficients now each coefficient becomes a power series in $(\epsilon,\mu)$.

The essential new feature is that now we also get a homological equation
for terms of degree 0 and 1 (previously these terms where absent by 
assumption). At degree 0 the homological equation states that the constant
vector must be in the kernel of the adjoint of $N$. The kernel of $N^*$ is 
$\hat{e}_3$, which gives the constant term in the unfolding. 
At degree 1 the homological equation states that the linear terms
must commute with the adjoint of $N$. The general matrix that 
commutes with $N^*$ is lower diagonal banded:
\[
\begin{pmatrix}
   \mu_3 & 0 & 0 \\
   \mu_2 & \mu_3 & 0 \\
   \mu_1 & \mu_2 & \mu_3 
\end{pmatrix} \,.
\]
Since we are dealing with divergence free vector fields we require $\mu_3 = 0$. 
Now the second component of the vector field would have an entry $x \mu_2$. To bring 
this into our simple form in which only the third component contains non-linear
and unfolding terms we can simply add an element from the range of the 
homological operator (at degree 1). Thus we can modify this by adding 
any matrix (times $\xi$) that does not commute with $N^*$. 
Hence we can simply remove $\hat{e}_2 x \mu_2$, since it is in the range.
In this way we find that the third component of the unfolded vector field becomes
\[
       \epsilon + \mu_1 x + \mu_2 y + p(x,y; \epsilon, \mu_1, \mu_2) \,.
\]
Under generic conditions on $p$ one of the two parameters $\mu_i$ can 
be removed. 
Crossing over from divergence free vector fields to volume preserving 
maps works as before. This proves theorem~3.

Interestingly the map thus obtained when $p$ is restricted to be a quadratic 
polynomials is exactly the map that was studied earlier in \cite{Lomeli03}.
In that paper this family of maps was studied because of its algebraic property that 
it is polynomial with polynomial inverse. 
We now see that this family of maps is also interesting because it is 
the unfolding of the volume preserving saddle-node bifurcation.

\section{Conclusion}\label{sec:conclusion}

While we have obtained a normal form for three dimensions, our results do not apply to the higher dimensional case. Indeed, \Tbl{tbl:dimtab}, which compares the dimension of the kernel of the homological operator in the volume-preserving subspace and the dimension of the set spanned by 
$\hat{e}_n  p(\xi_1, \dots, \xi_{n-1})$, shows that these two dimensions are no longer the same when $n>3$. It appears to be true that the dimension of the kernel is no larger and is strictly smaller for all degrees when $n>5$. Thus we would conjecture
that the span of $\hat{e}_n  p(\xi_1, \dots, \xi_{n-1})$ still contains a complement to the range so that
that our normal form can still be used (though it would not be the simplest normal form since it could contain more terms than strictly necessary). To prove this result the main problem is to obtain a description of the {\em polynomial} kernel of $\ad_N^{*}$; this becomes very complicated when $n> 3$, see \cite{Murdock03} for the case $n=4$. In higher dimensions most of the proof works, however, there is 
no simple general {\em polynomial} solution to the homological 
equation available as it was given in \eqref{eq:elphick3D}.

In a future paper we plan to study the dynamics of the unfolding of \eqref{eq:mapNF}.

\bibliographystyle{alpha}
\bibliography{VP111}

\end{document}